\documentclass[showpacs,aps,pra,twocolumn,superscriptaddress]{revtex4}

\usepackage{amsmath}
\usepackage{graphicx}
\usepackage{psfrag}

\newcommand{\un}[1]{\ensuremath{\ \mathrm{#1}}}
\newcommand{\sub}[1]{_{\mathrm{#1}}}
\renewcommand{\vec}[1]{\mathbf{#1}}

\begin{document}


\title{Deeply subrecoil two-dimensional Raman cooling}


\author{V. Boyer}
\affiliation{National Institute of Standards and Technology,
Gaithersburg, Maryland 20899, USA}
\affiliation{Clarendon Laboratory, University of Oxford, Oxford OX1
3PU, United Kingdom}

\author{L. J. Lising}
\altaffiliation[Permanent address: ]{Towson University Dept of
Physics, Astronomy, and Geosciences, Towson MD 21252, USA}
\affiliation{National Institute of Standards and Technology,
Gaithersburg, Maryland 20899, USA}

\author{S. L. Rolston}
\altaffiliation[Permanent address: ]{University of Maryland, College
Park, MD 20742, USA}
\affiliation{National Institute of Standards and Technology,
Gaithersburg, Maryland 20899, USA}

\author{W. D. Phillips}
\affiliation{National Institute of Standards and Technology,
Gaithersburg, Maryland 20899, USA}
\affiliation{Clarendon Laboratory, University of Oxford, Oxford OX1
3PU, United Kingdom}



\date{\today}

\begin{abstract}

We report the implementation of a two-dimensional Raman cooling scheme 
using sequential
excitations along the orthogonal axes. Using square pulses, we have
cooled a cloud of ultracold Cesium atoms
down to an RMS velocity spread of 0.39(5) recoil velocities,
corresponding to an effective temperature of 30\un{nK} (0.15~$T\sub{rec}$). This
technique can be useful to improve cold-atom atomic clocks, and is
particularly relevant for clocks in
microgravity.

\end{abstract}


\maketitle


\section{Introduction}

Types of laser cooling that involve atoms continually absorbing and emitting
photons cannot in general lead to atomic velocity distributions
narrower than the recoil velocity $v\sub{rec} = \hbar k/M$ where $k =
2\pi/\lambda$ is the wavevector of photons with wavelength $\lambda$
and $M$ is the atomic mass.  By contrast, Raman cooling~\cite{KaC92}
and velocity selective coherent population trapping~\cite{AAK88,
AAK89} can reach below this recoil limit.  These two techniques
involve an effective cessation of the absorption of the light by the
atoms once they have reached a sufficiently low velocity. We note
that the only application of subrecoil cooling of which
we are aware~\cite{BPR96} used one-dimensional (1D) Raman cooling.

Raman cooling has been demonstrated in one, two and three dimensions,
but deeply subrecoil velocities were only obtained in one
dimension~\cite{KaC92, RBD95}.  In two and three dimensions, the lowest
velocity spreads (1D RMS velocities) obtained were respectively
0.85~$v\sub{rec}$
and 1.34~$v\sub{rec}$~\cite{DLK94}. Defining the recoil temperature as
$k\sub{B} T\sub{rec} = M v\sub{rec}^2$, where $k\sub{B}$ is the 
Boltzmann constant, these correspond to 0.72 and 1.80~$T\sub{rec}$
respectively. In this
paper, we report the implementation of an efficient 2D Raman cooling
scheme that has produced velocity spreads as low as 0.39(5)~$v\sub{rec}$,
corresponding to 0.15~$T\sub{rec}$,
and that should, under appropriate circumstances, reach even lower
velocities. Our technique differs from that used previously in the
shape of the Raman pulses and the use of sequential excitations along
the orthogonal axes.

The use in atomic fountains of ultracold atoms,
produced by laser cooling in optical
molasses, has greatly improved the accuracy of
neutral-atom atomic clocks. Such
clocks work by launching the atoms vertically through a microwave
cavity, to which the atoms fall back after a Ramsey time as long as
about 1\un{s}. The opening in the microwave cavity has a diameter
typically less than 1\un{cm}, so the transverse temperature must be
low enough to allow a significant number of the launched atoms to pass
through the cavity the second time.  Atoms that do not make it through
the second time contribute to the collisional shift without
contributing to the signal. Fountain clocks experiments with
the coldest atoms achieve RMS spreads as low as
2~$v\sub{rec}$~\cite{JHD03}. Under these circumstances, many of the
atoms are clipped by the second passage through the cavity after 
1\un{s} of Ramsey time.
For significantly longer Ramsey times, for example as envisioned for
space-borne clocks, even lower temperatures, as obtained by subrecoil
cooling, will be needed.  Note that subrecoil longitudinal cooling is
not necessarily desirable, because the longitudinal thermal expansion
of the cloud reduces the atomic density and thus reduces the
collisional shift.  For that reason, the present work concentrates on
two-dimensional Raman cooling of an atomic sample released from
optical molasses, with a view to providing a valuable tool for future
atomic clocks.

The paper is organized as follows: In section~\ref{section1}, we
summarize Raman cooling theory. In section~\ref{section2}, we present
the experimental details, with a stress on the fine tuning of the
excitation spectrum of the Raman pulses.  Section~\ref{section3} gives
the results obtained with our apparatus, in terms of final velocity
distribution and cooling dynamics. In the conclusion, we summarize
and discuss our results and their applications.

\section{\label{section1} Raman cooling theory}

\begin{figure}[b]
  \includegraphics[width=\linewidth]{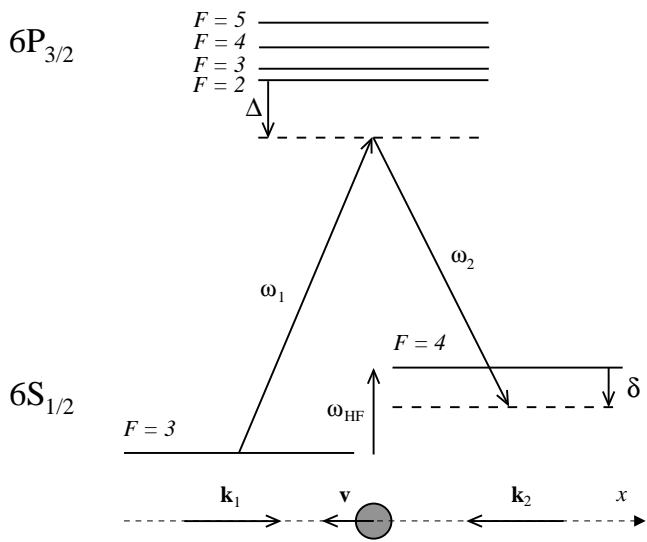}%
  \caption{\label{transition} On top, the Raman transition for the Cesium atom. At the bottom, the beam
  configuration for a velocity selective transition. We have shown the
  usual conditions where $\delta$ and $\Delta$ are negative.}
\end{figure}

The theory of one-dimensional Raman cooling in free space, described
in detail in~\cite{KaC92, MWK92}, is based on a two-step cycle. We
consider a cold cloud of cesium atoms initially in the hyperfine ground
state $6S_{1/2}, F=3$ (at this stage, we ignore the Zeeman, $ m_F$,
degeneracy).  First, the atoms are placed in the light field of two
off-resonant beams  counter-propagating along the $x$-axis, with
frequencies $\omega_1$ and $\omega_2$, and wavevectors $\vec{k}_1$ and
$\vec{k}_2$ such that $\vec{k}_1 \simeq -\vec{k}_2 \simeq \vec{k}$,
where $\vec{k}$ is a resonant wavevector for the transition $6S_{1/2}
\rightarrow 6P_{3/2}$ (Fig.~\ref{transition}). This light field
transfers the atoms with some (non zero) velocity along the beams to
the other hyperfine state $6S_{1/2}, F=4$, while changing their
velocity by two recoil velocities $\hbar(\vec{k}_1-\vec{k}_2)/M \simeq
2\hbar \vec{k}/M$.  The detuning $\Delta$ from the $6P_{3/2}$ manifold
is chosen to be much larger than the hyperfine splitting of the upper
state and also large enough to avoid one-photon excitation. Second, a
resonant pulse repumps the atoms to $F=3$ while giving them the
possibility to reach an $x$-velocity close to zero via the emission of
a spontaneous photon whose $x$-component of momentum can take any
value between $\pm \hbar k$. The Raman detuning $\delta$, shown in
Fig.~\ref{transition} and defined as $\delta = \omega_1 - \omega_2 -
\omega\sub{HF}$, is chosen to select atoms with velocities $\vec{v}$
fulfilling the resonance condition 
\begin{equation}
\delta = \delta\sub{LS}
 + \delta\sub{D} + 4\delta\sub{rec},
\label{resonance}
\end{equation}
where
\begin{eqnarray*}
\delta\sub{LS}&=&\frac{\Omega_1^2}{4}\left(\frac{1}{\Delta+\omega\sub{HF}} -
\frac{1}{\Delta} \right) -
\frac{\Omega_2^2}{4} \left(\frac{1}{\Delta-\omega\sub{HF}} -
\frac{1}{\Delta}\right),\\
\delta\sub{D} & = & 2\vec{k} \cdot \vec{v},\\
\delta\sub{rec} & = & \frac{E\sub{rec}}{\hbar} =  \frac{\hbar k^2}{2M}.
\end{eqnarray*}
In these equations, $\Omega_1$ and $\Omega_2$ are the effective
electric dipole couplings  of the Raman beams and $\omega\sub{HF}$ is
the hyperfine splitting. The three terms $\delta\sub{LS}$,
$\delta\sub{D}$ and $4\delta\sub{rec}$ are respectively the light
shifts, the Raman Doppler effect and four times the recoil energy shift.
This 3-level approach is a good approximation of our problem under the
following conditions: the polarizations of the Raman beams are linear
and the detuning $\Delta$ is large compared to the upper hyperfine
splitting (see the discussion in section~\ref{excitation_spectrum}).
Choosing $\delta < 0$ selects an initial velocity $\vec{v}$ whose
$x$-component is opposed to the velocity change, which is what we
want.  Cooling the opposite side of the velocity distribution implies
repeating the cycle with the directions of $\vec{k}_1$ and $\vec{k}_2$
reversed from that shown in Fig.~\ref{transition}.

Because of common mode rejection, only relative frequency noise
between the Raman laser beams affects the Raman selectivity. By phase
locking them relative to each other, this difference-frequency noise 
can be made much smaller than the
noise of their separate frequencies, and negligible. The excitation
spectrum is then fully determined by the shape and amplitude of the
pulses. A careful tailoring of the pulse shape allows a precise
excitation spectrum that does not excite atoms with a zero velocity
along the $x$-axis (see for example Fig.~\ref{spectrum}a). By repeating
the cooling cycle a large number of times, one forces the atoms to
perform a random walk in velocity space until they hit the zero
velocity state, a so-called dark state, where they tend to accumulate.

Our two-dimensional Raman cooling is a direct extension of the
one-dimensional case, where the cooling cycles are alternatively
applied to the $x$ and the $y$ directions.

The first Raman cooling experiments~\cite{KaC92, DLK94} used Blackman
pulses, which feature a power spectrum with very small wings outside
the central peak, hence reducing off-resonant excitations. Although
this might seem to be a very desirable feature, later
work~\cite{RBD95} showed experimentally and theoretically that square
pulses, which produce an excitation spectrum featuring significant
side lobes and a discrete set of zeros, give a better cooling in the
one-dimensional case. It is also expected to be better in the
two-dimensional case~\cite{RBD95, BBA01}. The dynamics of Raman
cooling, as well as that of VSCPT, are related to non Gaussian statistics
called L\'evy flights~\cite{BBA01, BBE94}. More precisely, for an
excitation spectrum varying as $v^\alpha$ around $v = 0$,
the width of the velocity
distribution scales with the cooling time $\Theta$ as $\Theta
^{-1/\alpha}$. However the atoms efficiently accumulate in the cold
peak of the distribution when $\Theta \rightarrow \infty$ only if
$\alpha$ is greater than or equal to the dimensionality of the
problem. Square pulses, for which $\alpha = 2$, appear to be
suitable for 2D cooling, and the present work concentrates on
them.

\section{\label{section2} Experimental setup}

\subsection{Laser system}

\begin{figure}[b]
  \includegraphics[width=.92\linewidth]{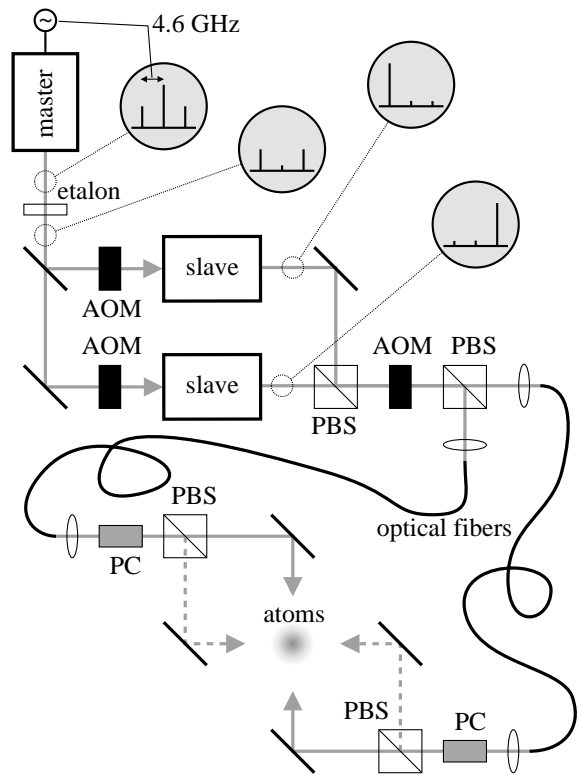}%
  \caption{\label{lasers} Schematic of the setup used to generate the Raman beams,
    showing the optical spectrum of the beams at various
    places. AOM: acousto-optic modulator; PBS: polarizing
    beam-splitter; PC: Pockels cell. The AOMs before the slave lasers
    are frequency shifters, and the AOM between the PBSs is a shutter.}
\end{figure}

Raman cooling requires two laser beams whose frequency difference is
locked to a frequency close to the hyperfine frequency
$\omega\sub{HF}/2\pi\simeq 9.2\un{GHz}$.  The most common methods used
to generate the two frequencies include direct electronic phase
locking of two free running lasers~\cite{SCL94}, acousto-optic
modulation~\cite{BGH93}, and electro-optic modulation~\cite{KaC92}. We
used a different approach based on current modulation of a laser
diode~\cite{RLG99, GTW83}, as shown in Fig.~\ref{lasers}. An
extended-cavity master diode laser at 852\un{nm}, with a free spectral range
of 4.6\un{GHz}, is current-modulated at $\omega\sub{HF}/2$ in order to
generate sidebands separated by $\omega\sub{HF}$. The fraction of the
power in the two first order sidebands, measured with an optical
spectrum analyzer, is about 50\%. The carrier is filtered out with a
solid etalon having a free spectral range of 9.2\un{GHz} and a finesse
of 8, and the remaining beam is used to injection-lock two slave
diodes. The slave currents are adjusted in order to lock one slave to
one sideband and the other slave to the other sideband. In the spectra
of the slaves, the total contamination from the carrier and any of the
unwanted sidebands is less than 1\% of the total power. The phase
coherence of the sidebands is fully transfered onto the slaves and the beatnote
spectrum of the two slaves is measured~\footnote{The linewidth of the beatnote
spectrum was measured by recording the beating of the two beams on a fast
photodiode. The photo-signal was mixed with the signal of an auxiliary
microwave generator tuned to a frequency close to 9.2\un{GHz}. The mixing
signal was analyzed with an FFT spectrum analyzer.} to be 1\un{Hz} wide. This
includes contributions from the linewidth of the microwave generator used to
modulate the master laser, the mechanical vibrations of the laser and
optical system (but
not of those mirrors after the fibers), and the resolution of the measurement
apparatus. After transport in optical fibers, 40\un{mW} are available in each
Raman beam.

As pointed out previously, cooling of opposite sides of the velocity
distribution requires interchanging the directions of $\vec{k}_1$ and
$\vec{k}_2$.
This
is done by interchanging the injection currents of the slaves so that the
sidebands to which they lock are interchanged, thus interchanging
their roles~\cite{SGC97}. The switching time, measured
by monitoring the transmission of each Raman beam through a confocal
cavity~\cite{SGC97}, is found to be about 30\un{\mu s} for a complete
switch, similar to what was observed in Ref.~\onlinecite{SGC97}. In the
cooling experiment, we allow an extra 20\un{\mu s} as a safety margin.
 The swapping of the beams between the $x$ and the
$y$ directions is done with Pockels cells and polarizing beam
splitters (Fig.~\ref{lasers}), in less than a microsecond. The
extinction ratio of the Pockels cell switches is about 100.

\subsection{Experimental details}

\begin{figure}
  \includegraphics[width=\linewidth]{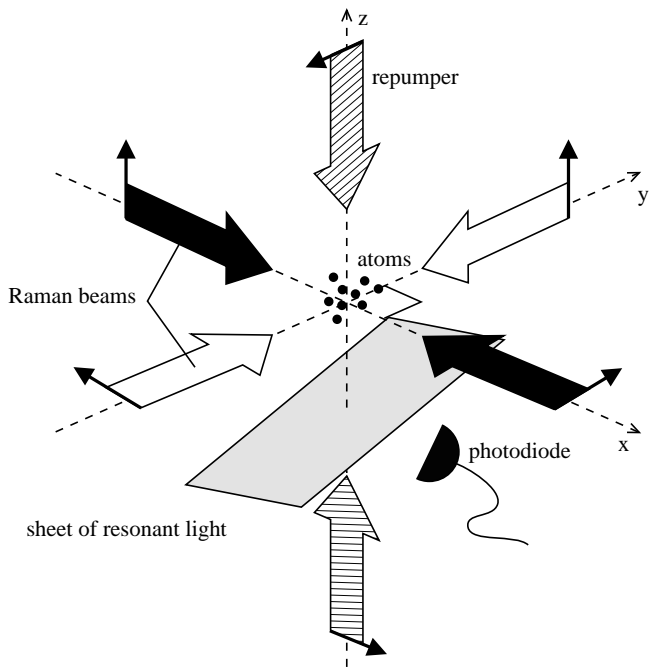}%
  \caption{\label{geometry} Geometry of the experiment. The Raman beams are in the
    horizontal plane, while the repumper beams are vertical. The
  polarizations are linear along the directions indicated by the
  arrows. The MOT beams are not shown. The atoms in $F=4$ are detected
  by fluorescence when they fall through a horizontal sheet of light 
  resonant
  with the transition $F=4 \rightarrow F'=5$, located 2\un{cm} under
  the MOT position.}
\end{figure}

A magneto-optical trap inside a glass cell is loaded with a few times
$10^7$ atoms from a chirped-slowed atomic beam. The cloud has an RMS
width of about 1~mm.  After additional, 70\un{ms}-long, molasses
cooling, the atoms are dropped, pumped into $F=3$, and Raman cooled
for 25\un{ms}, before they fall out of the Raman beams. As shown in
Fig.~\ref{geometry}, the Raman beams are in the horizontal plane,
along the $x$ and $y$ axes, providing cooling perpendicularly to the
vertical direction. We found that controlling the horizontality of the
Raman beams at a level of a few thousandths of a radian is enough to
ensure that gravity does not perturb the cooling, but an error as
large as 0.01 rad has a noticeable effect.  The waist of the beams
(radius at $1/e^2$ of peak intensity) is 4\un{mm} and they all have
nominally the same power. When the atoms are dropped, they are
slightly above the center of the beams, and after 25\un{ms} of
cooling, they are at an approximately symmetric position below.

The repumping is provided by a retro-reflected vertical beam, tuned to
the $F=4 \rightarrow F'=3$ transition, with an intensity a few times
the saturation intensity and with the reflected polarization rotated
in order to avoid a standing wave effect. The momentum of the
photons absorbed from the repumping beams has no effect on the
transverse velocity, and leads only to momentum diffusion in the
vertical direction.

The velocity distribution along $x$ or $y$ is measured by Raman
spectroscopy~\cite{KWR91}, \emph{i.e.} by transferring a narrow velocity class from $F = 3$
to $F = 4$ with a long Raman $\pi$ pulse.  Two centimeters below the
Raman beams, the atoms fall through a sheet of light tuned to the $F=4
\rightarrow F'=5$ cycling transition. The integrated fluorescence
collected by a photodiode is proportional to the number of atoms
transfered to the $F=4$ state.  By scanning the Raman detuning
$\delta$ for a succession of identically cooled atomic samples, one
can probe all the velocities and reconstruct the velocity
distribution. There is a small, uniform background signal; after
subtraction of this background, we obtain velocity distributions such
as that shown in Fig.~\ref{distribution}.

Raman cooling as described here is essentially a 3-level scheme and
our experiment requires the Zeeman sublevels to be degenerate within
each hyperfine level. Good subrecoil Raman cooling can be achieved
only if any Zeeman splitting is small compared to the Raman Doppler
shift associated with a single recoil velocity, which is
$8.2\un{kHz}$.  This is ensured by reducing the DC stray magnetic
field with an opposing applied external magnetic field, and further
reducing the DC and AC residual fields with a $\mu$-metal shield.
Raman spectroscopy with non-velocity-selective, co-propagating beams
and long pulses (300\un{\mu s}) is used to optimize the field zeroing
by adjusting for minimum spectral width. The Raman spectrum has a full
width half maximum (FWHM) of 0.5\un{kHz}, corresponding to a residual
stray field smaller than $100\un{\mu G}$, and equivalent to the Raman
Doppler shift of atoms with a velocity $v\sub{rec}/16$.

\subsection{\label{excitation_spectrum} Excitation spectrum}

The polarizations of each pair of Raman
beams are crossed-linear in order to ensure that, because the detuning
$|\Delta|$ is large compared to the hyperfine splitting 
(600\un{MHz}) of the excited
state, the light shifts are nearly the same for all the
Zeeman sublevels of the ground state~\cite{MCH93}. Under those
conditions, the effective electrical dipole coupling $\Omega$
corresponding to an intensity $I$ has a value $\Omega = \Gamma
\sqrt{0.67 \; I/2I_0}$, where $\Gamma = 2\pi \times 5.2\un{MHz}$ is
the natural linewidth of the excited state, and $I_0 = 1.1\un{mW/cm^2}$ is the
saturation intensity for the strongest transition. The Raman detuning
$\delta$ has to be negative to cool the atoms, and is chosen in such a way that
atoms with a zero velocity are resonant with the first zero point of the
excitation spectrum~\footnote{The Raman detuning $\delta$ is experimentally
adjusted (by optimizing the final velocity distribution) in such a way that the velocity class resonant with the first zero
point of the excitation spectrum is the same when we cool both sides of the
velocity distribution. However, because we do not know precisely the value of
the light shifts, nor do we have a perfect calibration of $\delta$, such a dark
state is not necessarily the zero velocity state. In fact, this degree of
freedom can be used to tune the direction of propagation of the atoms after the
cooling.}\label{note2.3}.  We extend Eq.~(\ref{resonance}) by defining the effective detuning
seen by these zero velocity atoms as
$$
\delta_0 = \delta - \delta\sub{LS} - 4\delta\sub{rec},
$$
so that
\begin{equation}
\delta_0 = \delta + 0.67 \frac{\Gamma^2
  I}{8 I_0} \left(\frac{1}{\Delta - \omega\sub{HF}} -
\frac{1}{\Delta + \omega\sub{HF}} \right) - \frac{2\hbar k^2}{M}.
\label{condition1}
\end{equation}
The excitation spectrum, defined as the probability $\mathcal{P}$ of
undergoing a Raman transition for any atom seeing an effective Raman
detuning $\delta\sub{ex}$, is given for a square pulse of length $t$
by the Rabi formula:
\begin{equation} 
\mathcal{P}(\delta\sub{ex}) = \frac{\Omega\sub{R}^2}{\delta\sub{ex}^2
  + \Omega\sub{R}^2} \sin^2 \left (\frac{t}{2} \sqrt{\delta\sub{ex}^2 +
  \Omega\sub{R}^2} \right),
\label{condition2}
\end{equation}
where $\Omega\sub{R} = \frac{I}{I_0} \cdot
\frac{\Gamma^2}{4|\Delta|}\mathcal{C}$ is the Raman Rabi frequency.
The coefficient $\mathcal{C}$ depends on the initial state in the
$F=3$ manifold. It has a mean value $\overline{\mathcal{C}}=0.28$ and
a total spread of $\pm20\%$.  The value of $\delta$ defined by
Eq.~(\ref{condition1}) must fulfill $\mathcal{P}(\delta_0) = 0$. That
is, for zero velocity atoms, $t \sqrt{\delta_0^2 + \Omega\sub{R}^2} =
2\pi$. In the above, we have assumed that $\Omega_1 = \Omega_2 =
\Omega$. In fact, the intensities $I_1$ and $I_2$ of the Raman beams
may differ by as much as 20\%. To take this into account, one would
have to write Eq.~(\ref{condition1}) in the form of
Eq.~(\ref{resonance}), and replace $I$ in the definition of
$\Omega\sub{R}$ with $\sqrt{I_1 I_2}$.

The detuning $\delta$ depends on the light intensity in two different
ways. Firstly, there is the differential light shift $\delta\sub{LS}$
between the two hyperfine levels, which is proportional to the light
intensity $I$.  Secondly, according to Eq.~(\ref{condition2}), the
frequency of the first zero of the excitation spectrum changes due to
a ``saturation'' effect, as soon as $\Omega\sub{R} t$ is not small
compared to 1, that is to say when the maximum transfer probability is
not small compared to unity. These dependences on the light intensity
lead to complications because the Raman beams have a Gaussian profile
and are never perfectly spatially homogeneous. Depending on their
position in the beams, different atoms have different resonance
conditions. In previous experiments~\cite{RBD95, Rei96}, the
differential light shift was reduced by using a detuning $\Delta$
large compare to the hyperfine splitting $\omega\sub{HF}$, and the
saturation effect was reduced by setting the maximum transfer
probability to 0.5, thus using $\pi/2$ pulses instead of $\pi$ pulses.

\begin{figure}[b]
  \includegraphics[width=0.92\linewidth]{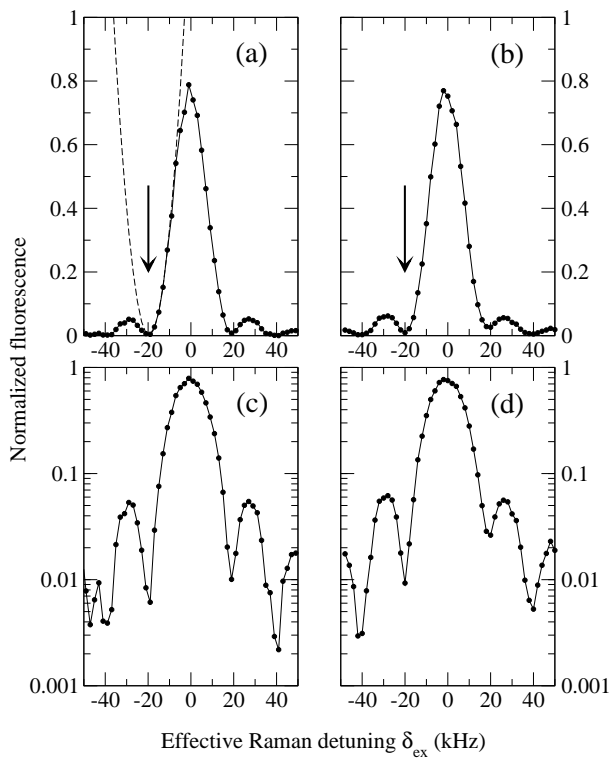}%
  \caption{\label{spectrum} Excitation spectrum of $50\un{\mu s}$-long Raman pulses,
    with a maximum transfer probability of 0.8. In figures~(a) and~(c),
    the atomic cloud is centered on the Raman beams (nearly homogeneous
    illumination),  in figures~(b) and~(d), 
    the cloud is on the side of the beams
    (inhomogeneous illumination). The position of the dark
    state is indicated by the arrows. The dashed line in figure~(a)
    is the parabola 
    that matches the second derivative of the spectrum at the position 
    of the dark state.}
\end{figure}

Because of a limitation in the available laser power, we worked at a
detuning $|\Delta|$ of only 20\un{GHz}, with pulses having a maximum
transfer efficiency of 80\% ($0.7\pi$-pulses). 
As shown below, although not negligible, the saturation shift
and the light shift partially cancel out, thus limiting unwanted
excitation of the dark state.

We measure the excitation spectrum by determining the transfer
efficiency of such Raman pulses as a function of the Raman detuning
$\delta$, with non-velocity-selective co-propagating beams.
Figure~\ref{spectrum} shows the excitation spectrum when the atomic
cloud is centered on the Raman beams (a), resulting in a fairly
homogeneous illumination, and when the cloud is on the edge of the
beams (b), resulting in an inhomogeneous illumination. The arrows show
the position of the Raman detuning corresponding to the dark state.
Because the light shift and the saturation respectively shift and
broaden the spectrum, locations in the cloud exposed to different
light intensities yield different positions and shapes of the
excitation spectrum. The spectra of the individual atoms contribute
inhomogeneously to the measured spectrum. In both case (a) and (b),
it appears that the spectrum is not fully symmetric and that the first
minimum on the positive detuning side does not go as close to zero as
does the minimum corresponding to the dark state. This comes from the
fact that the saturation and the light shift have opposite effects on
the position of the zero on one side of the spectrum, and similar
effects on the other side. This effect becomes very visible in the
inhomogeneous illumination case.  However, the cancellation is not
perfect, and the dispersion of the Raman coupling through the
dependence of $\mathcal{C}$ on the Zeeman sublevel also leads to a
``blurring'' of the spectrum.  As a result, the dark
state features a small excitation probability, even in the homogeneous
illumination case.  On the time scale of our cooling sequence, this has
not proven to be of importance.

With a detuning $|\Delta | = 20\un{GHz}$, our $50\un{\mu s}$-long 
Raman $0.7\pi$
pulse corresponds at the center of the beam to a mean Raman Rabi
frequency $\overline{\Omega}\sub{R} =
\overline{\mathcal{C}}\Gamma^2/4|\Delta| \cdot I / I_0 = 2\pi
\times 7\un{kHz}$, and an intensity $I\simeq 80\un{mW/cm^2}$ 
for each beam.  The probability of one-photon excitation is of the
order of $\Gamma /|\Delta|$ per pulse. It
results in a total probability of excitation of 10\% for a typical
cooling sequence of 300 pulses, which is low enough to avoid any
significant perturbation of the cooling.
The choice of a pulse length of 50\un{\mu s} means that the 
excitation spectrum covers
most of the initial velocity distribution, as seen on
Fig.~\ref{distribution}.

\section{\label{section3} Results}

The elementary Raman cooling cycle is a $50\un{\mu s}$ square Raman
pulse with the $\omega_1$ beam along some direction, e.g. $+x$ (as in
Fig.~\ref{transition}), and
the $\omega_2$ beam along the opposite direction, e.g.  $-x$, followed
by a $10\un{\mu s}$ resonant repumping pulse from counter-propagating
beams along $z$. In this example, the elementary cycle provides
cooling along the $+x$ axis. The rise and fall times of the Raman
pulse are less than $1\un{\mu s}$. In our experiment, the
elementary cooling cycles are applied in pairs along a given
direction. A complete cycle consists of 4 pairs of elementary cooling
cycles applied successively along the directions $+x$, $+y$, $-y$ and
$-x$. The full cooling sequence is typically composed of 40 complete
cycles.

As pointed out previously, the switching time between perpendicular
directions, limited by Pockels cell switching, is instantaneous with
respect to the experimental timescale, but the switching time between
parallel directions is $50\un{\mu s}$. Using pairs of identical
elementary cycles reduces the total number of complete cycles, and
thus reduces the total time spent switching the beams. There is almost
no loss of efficiency resulting from the application of two successive
elementary cycles along the same direction because after only a few
complete cooling cycles, the velocity distribution becomes narrow
enough so that the atoms are in the tail of the excitation spectrum
and  the excitation probability per Raman pulse is small compared to 1
for most of the atoms.

\begin{figure}[b]
  \makebox[0pt]{\makebox[18ex]{}\raisebox{-7ex}{\hspace*{2ex} \Large \textbf{(a)}}}%
  \includegraphics[angle=-90, width=\linewidth, totalheight=9cm, keepaspectratio=true]{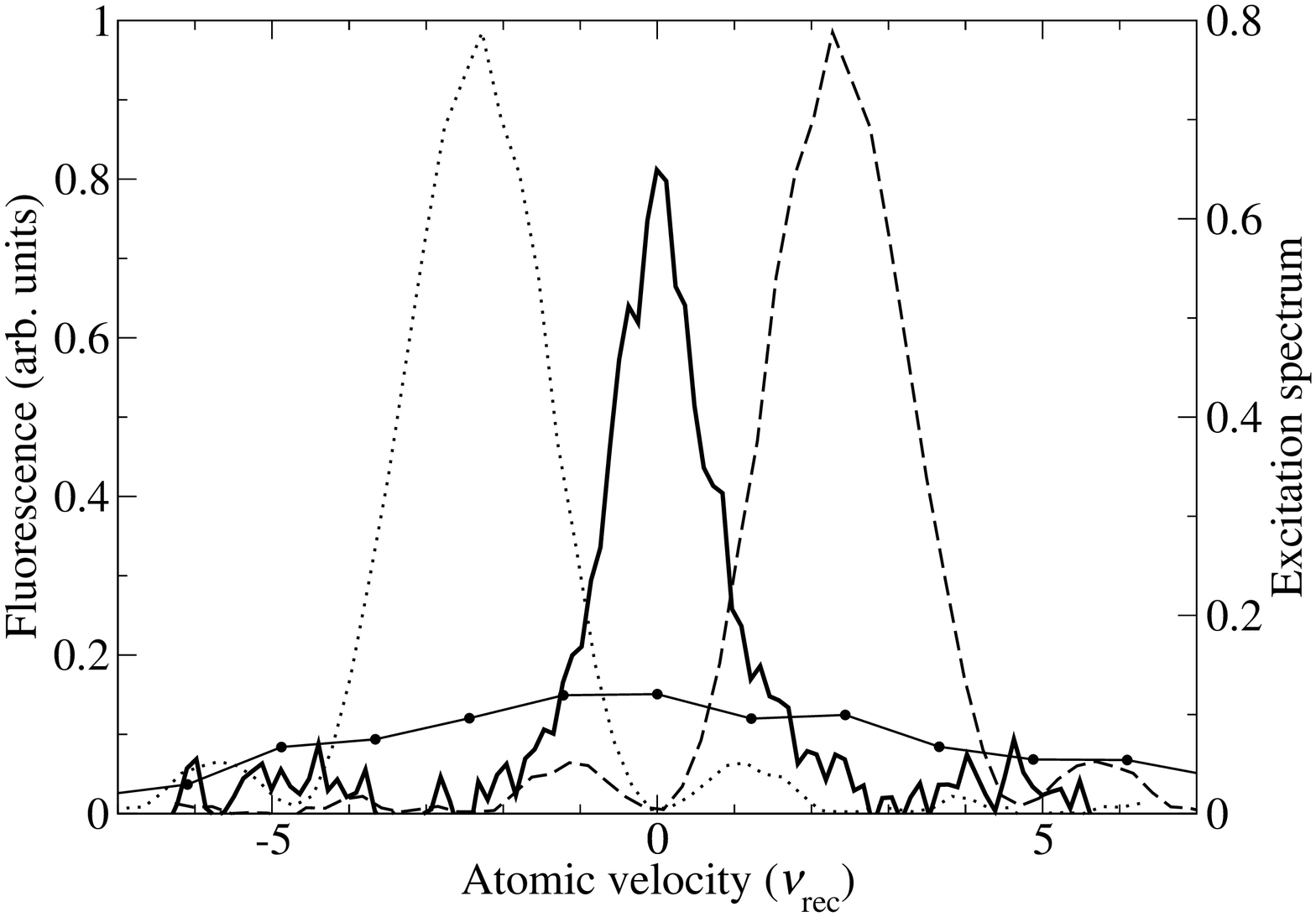}
  \makebox[0pt]{\makebox[18ex]{}\raisebox{-7ex}{\hspace*{2ex} \Large \textbf{(b)}}}%
  \includegraphics[angle=-90, width=\linewidth, totalheight=9cm, keepaspectratio=true]{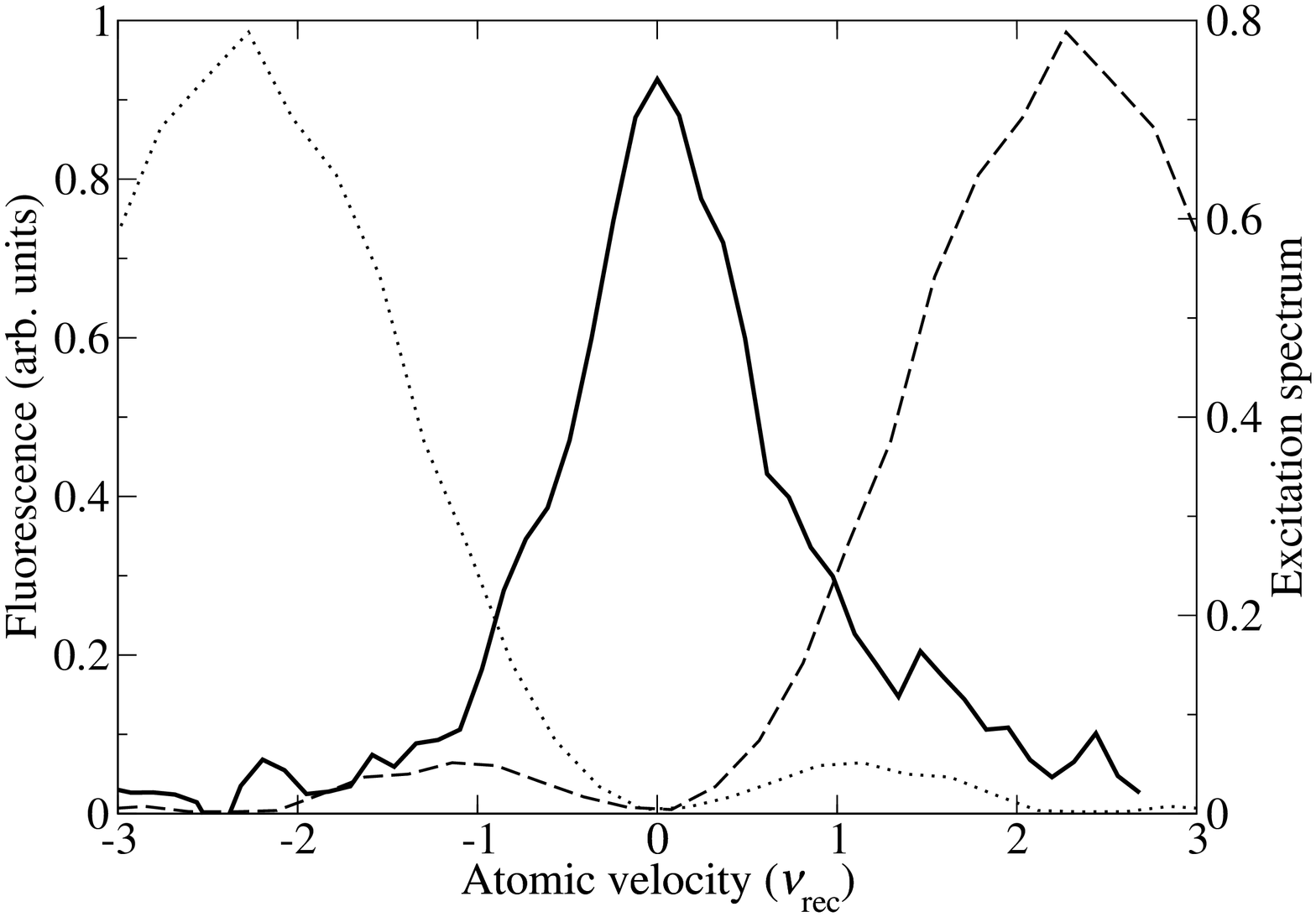}%
  \caption{\label{distribution}(a) Velocity distribution along the $x$ 
  axis before Raman
  cooling (circles), and after 15\un{ms} of 2D Raman cooling (heavy line).
  (b) Velocity distribution along the $x$ axis
  after a 22.4\un{ms} 2D Raman cooling 
  sequence consisting of 320 Raman pulses. Also
  shown in both graphs in dotted and dashed lines are the excitation spectra used to 
  cool along the directions $x$ and $-x$, tuned to match their
  first zeros on the dark state.
  }
\end{figure}

Immediately after the cooling, the velocity distribution is probed
with a $500\un{\mu s}$ square $\pi$ pulse applied along the $x$ or the
$y$ axis. Figure~\ref{distribution}b shows the velocity distribution
measured along the $x$ axis after 320 elementary cycles (40 complete
cycles), corresponding to a $22.4\un{ms}$ cooling sequence. The
velocity spread, defined as $\sigma = \mathrm{FWHM}/\sqrt{8 \ln 2}$,
is reduced from about 4~$v\sub{rec}$ after molasses cooling,
corresponding to an effective temperature of 3\un{\mu K}, to
0.39(5)~$v\sub{rec}$~\footnote{Uncertainties quoted in this paper
represent one standard deviation, combined statistical and systematic
uncertainties.}, corresponding to an effective temperature of
30\un{nK}. Note that a 20\% contribution from the probe excitation
linewidth is removed by deconvolution. A similar velocity
distribution is measured along the $y$ axis. 
For either of these 1D velocity distributions, about 80\% of the area
is contained in the central cold peak. Most of the rest of the atoms are
contained in two secondary peaks at positions where both excitation
spectra are close to zero, as seen in Fig.~\ref{distribution}a. From this we
conclude~\footnote{Simulations of the cooling process show that the
2D velocity distribution consists of one peak centered on the zero velocity
($v_x=0$, $v_y=0$), and 4 additional cold peaks
centered approximately at positions ($v_x=\pm
5v\sub{rec}$, $v_y=0$) and ($v_x=0$, $v_y=\pm 5v\sub{rec}$). These 4
peaks correspond to secondary dark
states and are experimentally determined to contain about 10\% of the atoms
each.} that in 2 dimensions, about 60\% of the atoms are in the 2D cold peak.

\begin{figure}
  \includegraphics[height=\linewidth, angle=-90]{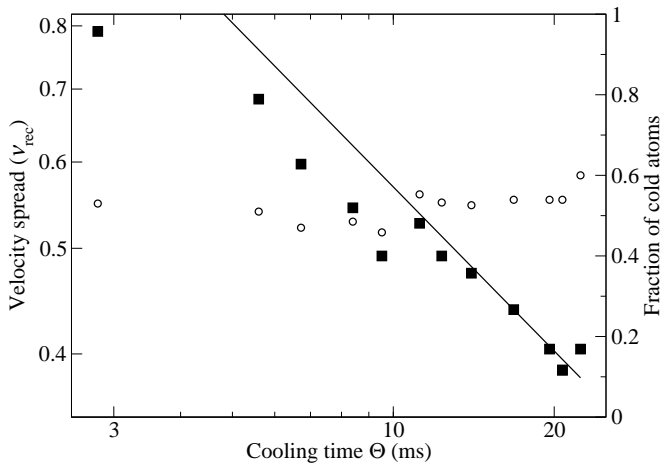}%
\caption{\label{dynamics} Filled squares: Velocity spread $\sigma$ as a function of the cooling time
  $\Theta$. The straight
  line is the power law
  $\beta \Theta^{-1/2}$, whose multiplicative factor $\beta$ has been
  adjusted manually to make the line fit the data at long times. Open
  circles: Fraction of atoms in the cold peak determined as explained
  in the text.}
\end{figure}

It is worth noting that because we operate in the case where
$\alpha$ is equal to the dimensionality of the cooling, the number of atoms
in the cold peak of the 2D velocity distribution is predicted~\cite{RBD95, BBA01} not to
change any
more as
soon as we enter the subrecoil regime and the
cold peak is clearly separated from the rest of the distribution. The main effect of continued
cooling is to reduce the width of the cold peak without changing the
number of atoms it contains. This is indeed what we observe as shown
in Fig.~\ref{dynamics}.

During the entire Raman cooling process, the 2D velocity width is reduced by a factor of 10,
while the size of the cloud along $x$ and $y$ does not change 
significantly, leading to a 100-fold increase of the 2D phase 
space density. Of course, taking into account the heating and
expansion along the $z$ direction (which we did not measure), the
increase in the 3D phase space density would be smaller, perhaps a few
tens.  We note however that Raman cooling in free space is not a
promising way to increase the phase space density to reach quantum
degeneracy because the spatial density stays roughly constant. This
means that a great deal of additional cooling is required.  On
the other hand, a high spatial density is not desirable for clock
applications, because it leads to collisional frequency shifts.

An important issue is the isotropy of the velocity distribution in the
cooling plane. The cooling scheme is fundamentally anisotropic, and we
only measured the velocity spread along the cooling directions $x$ and
$y$. However there are good reasons to believe that the peak of cold
atoms at the center of the final distribution is isotropic (in two
dimensions). As indicated in~\cite{BBE94}, the dynamics of each
individual atom is dominated by fairly distinct phases, where it
either performs a random walk in velocity space outside the subrecoil
range, or stays close to the dark state, in the subrecoil range, until
it gets excited and resumes the random walk. The argument for a final
isotropic velocity distribution relies on two considerations.
Firstly, the total excitation probability for an atom close to the
dark state during a complete cycle is isotropic. Indeed, close to the
null velocity, the excitation probabilities for an elementary cycle
along $\pm x$ or $\pm y$ are small compared to 1 and 
proportional respectively to $v_x^2$ and $v_y^2$ (power law with a
coefficient $\alpha = 2$),
leading to a total excitation probability for the complete cycle
proportional to the sum $v_r^2=v_x^2+v_y^2$. This probability only
depends on the ``distance'' $v_r$ from the dark state, and is therefore
isotropic.  Secondly, atoms excited from the subrecoil range perform
several steps during their random walk before going back to the
subrecoil range at a random point which is uncorrelated with the
previous position they occupied close to the dark state.
 
The combination of the effectively isotropic excitation of the
subrecoil atoms and the homogeneous filling of the subrecoil region
should produce an isotropic cold peak. We checked that a simple Monte
Carlo simulation ignoring Zeeman sublevels gives a perfectly isotropic
distribution in the $x-y$ plane.

We also studied the experimentally measured velocity spread $\sigma$
as a function of the cooling time $\Theta$. The results are shown on
Fig.~\ref{dynamics}.  As stated in section~\ref{section2}, subrecoil cooling
theory predicts that the velocity spread is described at long times by
the power law $\Theta^{-1/\alpha}$, where $\alpha$ is the excitation
spectrum power law coefficient.  Our data do not cover a range of
cooling times large enough to fully enter the asymptotic regime, and to allow
an accurate experimental determination of $\alpha$. However,
Fig.~\ref{dynamics} shows that our data is compatible with cooling
dynamics described by the theoretical power law $\Theta^{-1/2}$
($\alpha=2$) at times longer than 10\un{ms}.

It is expected that because of the experimental imperfections, the
velocity spread would eventually reach a finite value at long times.
Nonetheless, Fig.~\ref{dynamics} does not show any evident
saturation, which indicates that a longer cooling time, for
instance in micro-gravity, would lead to an even smaller velocity spread.

One of the potential limitations of Raman cooling is photon
reabsorption~\cite{CCL98}.
However, with a spatial density less than $10^{10}\un{cm^{-3}}$ and an optical
depth of the order of unity, we do not expect photon
reabsorption to hinder the cooling~\cite{PKB99}. In clock 
applications, the density would typically be significantly 
lower than in these experiments and reabsorption should be completely 
negligible.

To keep the cooling sequence simple, we use Raman
pulses with a fixed length. The choice of 50\un{\mu s}-long pulses is
convenient because the resulting excitation spectrum covers most of
the initial velocity distribution, and more importantly, covers the
maximum excursion range of the atoms during their random walk. Indeed,
an atom close to zero velocity can be pushed away from the center
of the velocity distribution by a maximum amount of about
4~$v\sub{rec}$: two recoil velocities during the Raman transition plus
one or two (or exceptionally more) during the repumping process. The
excitation spectrum covers 5~$v\sub{rec}$ between the two first zeros.  

It should be possible to use longer pulses (narrower excitation
spectrum) in combination with short pulses (wider excitation
spectrum) that recycle atoms far from zero velocity~\cite{RBD95}.
The increased filtering effect of longer pulses produces an narrower
distribution at the cost of a smaller number of atoms in the cold
peak. In any case, the best cooling strategy results from a tradeoff
between the width of the distribution and the fraction of atoms in the
cold peak, and depends on the total cooling time available.

\section{Conclusions}

Our scheme produces a narrower velocity distribution with respect to
the recoil velocity than what was previously achieved with 2D Raman
cooling~\cite{DLK94}. There are two main differences from that
previous scheme, where the cooling was performed in a vertical plane,
from the four directions at the same time.  First, we use one pair of
Raman beams at a time, in order to avoid unwanted excitation of the
dark state. Indeed, in Ref.~\cite{DLK94}, the careful use of circular
polarization avoided diffracting the atoms from standing waves created
by counter-propagating beams of the same frequency, but higher order
photon transitions of the type $(\omega_1, \vec{k})
(\omega_2,\vec{-k}) (\omega_2,\vec{k}) (\omega_1,\vec{-k})$ would still
be able to transfer 4~$v\sub{rec}$ to the atoms.  Second,
since we cool in the horizontal plane, gravity has no effect on the
velocity components which are being cooled and does not accelerate
atoms out of the cold peak, allowing for a more effective cooling.

To be used in an cold-atom atomic clock, Raman cooling has to be
coupled with a launching mechanism like moving molasses. An easy
solution is to first launch the atoms and then collimate them with
Raman cooling on their way towards the first microwave cavity. The
cooling time depends on the size of the Raman beams and the launch
speed. 

In that perspective, our setup performs quite well in comparison with
a newer scheme relying on sideband cooling in optical
lattices~\cite{HHK98}, which has produced polarized samples with a
velocity spread of 0.85~$v\sub{rec}$ in 3D, in a fountain-like
geometry~\cite{TCC01}.  While it is appealing for its simplicity,
sideband cooling has a fundamental limit for the lowest velocity
spread achievable, which is about 0.7 times the recoil velocity
associated with the wavevector of the lattice used to trap the
atoms~\cite{KPR95}. Raman cooling has no such a limitation.

It is worth noticing that, although deeply subrecoil velocities are
obtained for long cooling times, only accessible in micro-gravity, 2D
Raman cooling can still provide subrecoil velocities in a few
milliseconds, as shown in Fig.~\ref{dynamics}. Implementing the scheme
on a moving-molasses earth-bound fountain, where a
1.5\un{cm}-interaction region with the Raman beams combined with a
typical launch velocity of 5\un{m/s} leads to an interaction time of
3\un{ms}, would yield a substantial improvement in terms of
brightness of the atomic source, reducing the transverse velocity
spread from a few recoils to less than a recoil velocity.

For a micro-gravity-operated atomic beam, the improvement would be even
more dramatic because the launch velocity can be much smaller than in
a fountain, making the interaction time with the Raman beams much
longer. The maximum cooling time is more likely to be limited by the
maximum longitudinal heating acceptable. How the increase of
brightness translates into an increase of the stability of a
space-borne atomic clock depends on geometrical details and on the
factors which actually limit the stability and/or the accuracy. For
simplicity, let us assume that the atomic cloud is severely clipped by
the opening of the second microwave cavity, as is the case in
current fountain clocks, and that the signal-to-noise ratio is the
main limiting factor of the short-term stability~\footnote{One could
choose instead to take advantage of the improved collimation by
reducing the initial number of atoms launched in order to reduce the
collisional shift, which is a major source of inaccuracy in laser
cooled Cesium
atomic clocks. However, recent developments~\cite{PMB02} suggest
that the collisional shift can be accurately measured and accounted
for.}. Reducing the transverse velocity spread from 2~$v\sub{rec}$
to 0.4~$v\sub{rec}$ (our
current result) would increase the flux of atoms through the cavity by
a factor of 25.  That would translate into a 5-fold increase of the
signal-to-noise ratio, leading to a 5-fold increase of the short-term
stability for a given averaging time, or a 25-fold reduction in the
averaging time needed to achieve a given short-term stability.

\begin{acknowledgments}
We thank F. Bardou and C. Ekstrom for very helpful discussions.
We also thank C. Ekstrom, W.M. Golding and S. Ghezali for early 
contributions to the experimental apparatus.
This work was funded in part by ONR and NASA. 
\end{acknowledgments}

\bibliography{article.bib}

\begin{thebibliography}{24}
\expandafter\ifx\csname natexlab\endcsname\relax\def\natexlab#1{#1}\fi
\expandafter\ifx\csname bibnamefont\endcsname\relax
  \def\bibnamefont#1{#1}\fi
\expandafter\ifx\csname bibfnamefont\endcsname\relax
  \def\bibfnamefont#1{#1}\fi
\expandafter\ifx\csname citenamefont\endcsname\relax
  \def\citenamefont#1{#1}\fi
\expandafter\ifx\csname url\endcsname\relax
  \def\url#1{\texttt{#1}}\fi
\expandafter\ifx\csname urlprefix\endcsname\relax\def\urlprefix{URL }\fi
\providecommand{\bibinfo}[2]{#2}
\providecommand{\eprint}[2][]{\url{#2}}

\bibitem[{\citenamefont{Kasevich and Chu}(1992)}]{KaC92}
\bibinfo{author}{\bibfnamefont{M.}~\bibnamefont{Kasevich}} \bibnamefont{and}
  \bibinfo{author}{\bibfnamefont{S.}~\bibnamefont{Chu}},
  \bibinfo{journal}{Phys. Rev. Lett.} \textbf{\bibinfo{volume}{69}},
  \bibinfo{pages}{1741} (\bibinfo{year}{1992}).

\bibitem[{\citenamefont{Aspect et~al.}(1988)\citenamefont{Aspect, Arimondo,
  Kaiser, Vansteenkiste, and Cohen-Tannoudji}}]{AAK88}
\bibinfo{author}{\bibfnamefont{A.}~\bibnamefont{Aspect}},
  \bibinfo{author}{\bibfnamefont{E.}~\bibnamefont{Arimondo}},
  \bibinfo{author}{\bibfnamefont{R.}~\bibnamefont{Kaiser}},
  \bibinfo{author}{\bibfnamefont{N.}~\bibnamefont{Vansteenkiste}},
  \bibnamefont{and}
  \bibinfo{author}{\bibfnamefont{C.}~\bibnamefont{Cohen-Tannoudji}},
  \bibinfo{journal}{Phys. Rev. Lett.} \textbf{\bibinfo{volume}{61}},
  \bibinfo{pages}{826} (\bibinfo{year}{1988}).

\bibitem[{\citenamefont{Aspect et~al.}(1989)\citenamefont{Aspect, Arimondo,
  Kaiser, Vansteenkiste, and Cohen-Tannoudji}}]{AAK89}
\bibinfo{author}{\bibfnamefont{A.}~\bibnamefont{Aspect}},
  \bibinfo{author}{\bibfnamefont{E.}~\bibnamefont{Arimondo}},
  \bibinfo{author}{\bibfnamefont{R.}~\bibnamefont{Kaiser}},
  \bibinfo{author}{\bibfnamefont{N.}~\bibnamefont{Vansteenkiste}},
  \bibnamefont{and}
  \bibinfo{author}{\bibfnamefont{C.}~\bibnamefont{Cohen-Tannoudji}},
  \bibinfo{journal}{J. Opt. Soc. Am.} \textbf{\bibinfo{volume}{6}},
  \bibinfo{pages}{2112} (\bibinfo{year}{1989}).

\bibitem[{\citenamefont{BenDahan et~al.}(1996)\citenamefont{BenDahan, Peik,
  Reichel, Castin, and Salomon}}]{BPR96}
\bibinfo{author}{\bibfnamefont{M.}~\bibnamefont{BenDahan}},
  \bibinfo{author}{\bibfnamefont{E.}~\bibnamefont{Peik}},
  \bibinfo{author}{\bibfnamefont{J.}~\bibnamefont{Reichel}},
  \bibinfo{author}{\bibfnamefont{Y.}~\bibnamefont{Castin}}, \bibnamefont{and}
  \bibinfo{author}{\bibfnamefont{C.}~\bibnamefont{Salomon}},
  \bibinfo{journal}{Phys. Rev. Lett.} \textbf{\bibinfo{volume}{76}},
  \bibinfo{pages}{4508} (\bibinfo{year}{1996}).

\bibitem[{\citenamefont{Reichel et~al.}(1995)}]{RBD95}
\bibinfo{author}{\bibfnamefont{J.}~\bibnamefont{Reichel}} \bibnamefont{et~al.},
  \bibinfo{journal}{Phys. Rev. Lett.} \textbf{\bibinfo{volume}{75}},
  \bibinfo{pages}{4575} (\bibinfo{year}{1995}).

\bibitem[{\citenamefont{Davidson et~al.}(1994)\citenamefont{Davidson, Lee,
  Kasevich, and Chu}}]{DLK94}
\bibinfo{author}{\bibfnamefont{N.}~\bibnamefont{Davidson}},
  \bibinfo{author}{\bibfnamefont{H.-J.} \bibnamefont{Lee}},
  \bibinfo{author}{\bibfnamefont{M.}~\bibnamefont{Kasevich}}, \bibnamefont{and}
  \bibinfo{author}{\bibfnamefont{S.}~\bibnamefont{Chu}},
  \bibinfo{journal}{Phys. Rev. Lett.} \textbf{\bibinfo{volume}{72}},
  \bibinfo{pages}{3158} (\bibinfo{year}{1994}), \bibinfo{note}{note that this
  reference quotes 2D and 3D RMS velocities which we converted to 1D RMS
  velocities by dividing by $\sqrt{2}$ and $\sqrt{3}$ respectively. Note also
  that our definition of $T\sub{rec}$ is different from that of Davidson {\em
  et al}.}

\bibitem[{\citenamefont{Jefferts et~al.}(2003)\citenamefont{Jefferts, Heavner,
  Donley, Shirley, and Parker}}]{JHD03}
\bibinfo{author}{\bibfnamefont{S.}~\bibnamefont{Jefferts}},
  \bibinfo{author}{\bibfnamefont{T.}~\bibnamefont{Heavner}},
  \bibinfo{author}{\bibfnamefont{E.}~\bibnamefont{Donley}},
  \bibinfo{author}{\bibfnamefont{J.}~\bibnamefont{Shirley}}, \bibnamefont{and}
  \bibinfo{author}{\bibfnamefont{T.}~\bibnamefont{Parker}}, in
  \emph{\bibinfo{booktitle}{Proc. 2003 Joint Mtg. IEEE Intl. Freq. Cont. Symp.
  and EFTF Conf.}} (\bibinfo{year}{2003}), p. \bibinfo{pages}{1084}.

\bibitem[{\citenamefont{Moler et~al.}(1992)\citenamefont{Moler, Weiss,
  Kasevich, and Chu}}]{MWK92}
\bibinfo{author}{\bibfnamefont{K.}~\bibnamefont{Moler}},
  \bibinfo{author}{\bibfnamefont{D.~S.} \bibnamefont{Weiss}},
  \bibinfo{author}{\bibfnamefont{M.}~\bibnamefont{Kasevich}}, \bibnamefont{and}
  \bibinfo{author}{\bibfnamefont{S.}~\bibnamefont{Chu}},
  \bibinfo{journal}{Phys. Rev. A} \textbf{\bibinfo{volume}{45}},
  \bibinfo{pages}{342} (\bibinfo{year}{1992}).

\bibitem[{\citenamefont{Bardou et~al.}(2001)\citenamefont{Bardou, Bouchaud,
  Aspect, and Cohen-Tannoudji}}]{BBA01}
\bibinfo{author}{\bibfnamefont{F.}~\bibnamefont{Bardou}},
  \bibinfo{author}{\bibfnamefont{J.-P.} \bibnamefont{Bouchaud}},
  \bibinfo{author}{\bibfnamefont{A.}~\bibnamefont{Aspect}}, \bibnamefont{and}
  \bibinfo{author}{\bibfnamefont{C.}~\bibnamefont{Cohen-Tannoudji}},
  \emph{\bibinfo{title}{L\'evy statistics and laser cooling}}
  (\bibinfo{publisher}{Cambridge University Press}, \bibinfo{year}{2001}).

\bibitem[{\citenamefont{Bardou et~al.}(1994)\citenamefont{Bardou, Bouchaud,
  Emile, Aspect, and Cohen-Tannoudji}}]{BBE94}
\bibinfo{author}{\bibfnamefont{F.}~\bibnamefont{Bardou}},
  \bibinfo{author}{\bibfnamefont{J.~P.} \bibnamefont{Bouchaud}},
  \bibinfo{author}{\bibfnamefont{O.}~\bibnamefont{Emile}},
  \bibinfo{author}{\bibfnamefont{A.}~\bibnamefont{Aspect}}, \bibnamefont{and}
  \bibinfo{author}{\bibfnamefont{C.}~\bibnamefont{Cohen-Tannoudji}},
  \bibinfo{journal}{Phys. Rev. Lett.} \textbf{\bibinfo{volume}{72}},
  \bibinfo{pages}{203} (\bibinfo{year}{1994}).

\bibitem[{\citenamefont{Santarelli et~al.}(1994)\citenamefont{Santarelli,
  Clairon, Lea, and Tino}}]{SCL94}
\bibinfo{author}{\bibfnamefont{G.}~\bibnamefont{Santarelli}},
  \bibinfo{author}{\bibfnamefont{A.}~\bibnamefont{Clairon}},
  \bibinfo{author}{\bibfnamefont{S.~N.} \bibnamefont{Lea}}, \bibnamefont{and}
  \bibinfo{author}{\bibfnamefont{G.}~\bibnamefont{Tino}},
  \bibinfo{journal}{Opt. Commun.} \textbf{\bibinfo{volume}{104}},
  \bibinfo{pages}{339} (\bibinfo{year}{1994}).

\bibitem[{\citenamefont{Bouyer et~al.}(1993)\citenamefont{Bouyer, Gustavson,
  Haritos, and Kasevich}}]{BGH93}
\bibinfo{author}{\bibfnamefont{P.}~\bibnamefont{Bouyer}},
  \bibinfo{author}{\bibfnamefont{T.~L.} \bibnamefont{Gustavson}},
  \bibinfo{author}{\bibfnamefont{K.~G.} \bibnamefont{Haritos}},
  \bibnamefont{and} \bibinfo{author}{\bibfnamefont{M.~A.}
  \bibnamefont{Kasevich}}, \bibinfo{journal}{Opt. Lett.}
  \textbf{\bibinfo{volume}{18}}, \bibinfo{pages}{649} (\bibinfo{year}{1993}).

\bibitem[{\citenamefont{Ringot et~al.}(1999)\citenamefont{Ringot, Lecoq,
  Garreau, and Szriftgiser}}]{RLG99}
\bibinfo{author}{\bibfnamefont{J.}~\bibnamefont{Ringot}},
  \bibinfo{author}{\bibfnamefont{Y.}~\bibnamefont{Lecoq}},
  \bibinfo{author}{\bibfnamefont{J.~C.} \bibnamefont{Garreau}},
  \bibnamefont{and}
  \bibinfo{author}{\bibfnamefont{P.}~\bibnamefont{Szriftgiser}},
  \bibinfo{journal}{Eur. Phys. J. D} \textbf{\bibinfo{volume}{7}},
  \bibinfo{pages}{285} (\bibinfo{year}{1999}).

\bibitem[{\citenamefont{Goldberg et~al.}(1983)\citenamefont{Goldberg, Taylor,
  Weller, and Boom}}]{GTW83}
\bibinfo{author}{\bibfnamefont{L.}~\bibnamefont{Goldberg}},
  \bibinfo{author}{\bibfnamefont{H.~F.} \bibnamefont{Taylor}},
  \bibinfo{author}{\bibfnamefont{J.~F.} \bibnamefont{Weller}},
  \bibnamefont{and} \bibinfo{author}{\bibfnamefont{D.}~\bibnamefont{Boom}},
  \bibinfo{journal}{Electron. Lett.} \textbf{\bibinfo{volume}{19}},
  \bibinfo{pages}{491} (\bibinfo{year}{1983}).

\bibitem[{\citenamefont{Szymaniec et~al.}(1997)\citenamefont{Szymaniec,
  Ghezali, Cognet, and Clairon}}]{SGC97}
\bibinfo{author}{\bibfnamefont{K.}~\bibnamefont{Szymaniec}},
  \bibinfo{author}{\bibfnamefont{S.}~\bibnamefont{Ghezali}},
  \bibinfo{author}{\bibfnamefont{L.}~\bibnamefont{Cognet}}, \bibnamefont{and}
  \bibinfo{author}{\bibfnamefont{A.}~\bibnamefont{Clairon}},
  \bibinfo{journal}{Opt. Commun.} \textbf{\bibinfo{volume}{144}},
  \bibinfo{pages}{50} (\bibinfo{year}{1997}).

\bibitem[{\citenamefont{Kasevich et~al.}(1991)\citenamefont{Kasevich, Weiss,
  Riis, Moler, Kasapi, and Chu}}]{KWR91}
\bibinfo{author}{\bibfnamefont{M.}~\bibnamefont{Kasevich}},
  \bibinfo{author}{\bibfnamefont{D.~S.} \bibnamefont{Weiss}},
  \bibinfo{author}{\bibfnamefont{E.}~\bibnamefont{Riis}},
  \bibinfo{author}{\bibfnamefont{K.}~\bibnamefont{Moler}},
  \bibinfo{author}{\bibfnamefont{S.}~\bibnamefont{Kasapi}}, \bibnamefont{and}
  \bibinfo{author}{\bibfnamefont{S.}~\bibnamefont{Chu}},
  \bibinfo{journal}{Phys. Rev. Lett.} \textbf{\bibinfo{volume}{66}},
  \bibinfo{pages}{2297} (\bibinfo{year}{1991}).

\bibitem[{\citenamefont{Miller et~al.}(1993)\citenamefont{Miller, Cline, and
  Heinzen}}]{MCH93}
\bibinfo{author}{\bibfnamefont{J.~D.} \bibnamefont{Miller}},
  \bibinfo{author}{\bibfnamefont{R.~A.} \bibnamefont{Cline}}, \bibnamefont{and}
  \bibinfo{author}{\bibfnamefont{D.~J.} \bibnamefont{Heinzen}},
  \bibinfo{journal}{Phys. Rev. A} \textbf{\bibinfo{volume}{47}},
  \bibinfo{pages}{R4567} (\bibinfo{year}{1993}).

\bibitem[{\citenamefont{Reichel}(1996)}]{Rei96}
\bibinfo{author}{\bibfnamefont{J.}~\bibnamefont{Reichel}}, Ph.D. thesis,
  \bibinfo{school}{University of Paris VI} (\bibinfo{year}{1996}).

\bibitem[{\citenamefont{Castin et~al.}(1998)\citenamefont{Castin, Cirac, and
  Lewenstein}}]{CCL98}
\bibinfo{author}{\bibfnamefont{Y.}~\bibnamefont{Castin}},
  \bibinfo{author}{\bibfnamefont{J.~I.} \bibnamefont{Cirac}}, \bibnamefont{and}
  \bibinfo{author}{\bibfnamefont{M.}~\bibnamefont{Lewenstein}},
  \bibinfo{journal}{Phys. Rev. Lett.} \textbf{\bibinfo{volume}{80}},
  \bibinfo{pages}{5305} (\bibinfo{year}{1998}).

\bibitem[{\citenamefont{Perrin et~al.}(1999)\citenamefont{Perrin, Kuhn,
  Bouchoule, Pfau, and Salomon}}]{PKB99}
\bibinfo{author}{\bibfnamefont{H.}~\bibnamefont{Perrin}},
  \bibinfo{author}{\bibfnamefont{A.}~\bibnamefont{Kuhn}},
  \bibinfo{author}{\bibfnamefont{I.}~\bibnamefont{Bouchoule}},
  \bibinfo{author}{\bibfnamefont{T.}~\bibnamefont{Pfau}}, \bibnamefont{and}
  \bibinfo{author}{\bibfnamefont{C.}~\bibnamefont{Salomon}},
  \bibinfo{journal}{Europhys. Lett.} \textbf{\bibinfo{volume}{46}},
  \bibinfo{pages}{141} (\bibinfo{year}{1999}).

\bibitem[{\citenamefont{Hamann et~al.}(1998)\citenamefont{Hamann, Haycock,
  Klose, Pax, Deutsch, and Jessen}}]{HHK98}
\bibinfo{author}{\bibfnamefont{S.~E.} \bibnamefont{Hamann}},
  \bibinfo{author}{\bibfnamefont{D.~L.} \bibnamefont{Haycock}},
  \bibinfo{author}{\bibfnamefont{G.}~\bibnamefont{Klose}},
  \bibinfo{author}{\bibfnamefont{P.~H.} \bibnamefont{Pax}},
  \bibinfo{author}{\bibfnamefont{I.~H.} \bibnamefont{Deutsch}},
  \bibnamefont{and} \bibinfo{author}{\bibfnamefont{P.~S.}
  \bibnamefont{Jessen}}, \bibinfo{journal}{Phys. Rev. Lett.}
  \textbf{\bibinfo{volume}{80}}, \bibinfo{pages}{4149} (\bibinfo{year}{1998}).

\bibitem[{\citenamefont{Treutlein et~al.}(2001)\citenamefont{Treutlein, Chung,
  and Chu}}]{TCC01}
\bibinfo{author}{\bibfnamefont{P.}~\bibnamefont{Treutlein}},
  \bibinfo{author}{\bibfnamefont{K.~Y.} \bibnamefont{Chung}}, \bibnamefont{and}
  \bibinfo{author}{\bibfnamefont{S.}~\bibnamefont{Chu}},
  \bibinfo{journal}{Phys. Rev. A} \textbf{\bibinfo{volume}{63}},
  \bibinfo{pages}{051401} (\bibinfo{year}{2001}).

\bibitem[{\citenamefont{Kastberg et~al.}(1995)\citenamefont{Kastberg, Phillips,
  Rolston, Spreeuw, and Jessen}}]{KPR95}
\bibinfo{author}{\bibfnamefont{A.}~\bibnamefont{Kastberg}},
  \bibinfo{author}{\bibfnamefont{W.~D.} \bibnamefont{Phillips}},
  \bibinfo{author}{\bibfnamefont{S.~L.} \bibnamefont{Rolston}},
  \bibinfo{author}{\bibfnamefont{R.~J.~C.} \bibnamefont{Spreeuw}},
  \bibnamefont{and} \bibinfo{author}{\bibfnamefont{P.~S.}
  \bibnamefont{Jessen}}, \bibinfo{journal}{Phys. Rev. Lett.}
  \textbf{\bibinfo{volume}{74}}, \bibinfo{pages}{1542} (\bibinfo{year}{1995}).

\bibitem[{\citenamefont{DosSantos et~al.}(2002)\citenamefont{DosSantos, Marion,
  Bize, Sortais, Clairon, and Salomon}}]{PMB02}
\bibinfo{author}{\bibfnamefont{F.~P.} \bibnamefont{DosSantos}},
  \bibinfo{author}{\bibfnamefont{H.}~\bibnamefont{Marion}},
  \bibinfo{author}{\bibfnamefont{S.}~\bibnamefont{Bize}},
  \bibinfo{author}{\bibfnamefont{Y.}~\bibnamefont{Sortais}},
  \bibinfo{author}{\bibfnamefont{A.}~\bibnamefont{Clairon}}, \bibnamefont{and}
  \bibinfo{author}{\bibfnamefont{C.}~\bibnamefont{Salomon}},
  \bibinfo{journal}{Phys. Rev. Lett.} \textbf{\bibinfo{volume}{89}},
  \bibinfo{pages}{233004} (\bibinfo{year}{2002}).

\end{thebibliography}

\end{document}